\documentclass[superscriptaddress,pre]{revtex4}

\usepackage{epsfig}

\begin{document}

\title{Forces between functionalized silica nanoparticles in solution}

\author{J. Matthew D. Lane}
\author{Ahmed E. Ismail}
\author{Michael Chandross}
\affiliation{Sandia National Laboratories, Albuquerque, NM 87185}

\author{Christian D. Lorenz}
\affiliation{Materials Research Group, King's College, London, WC2R 2LS, UK}

\author{Gary S. Grest}
\affiliation{Sandia National Laboratories, Albuquerque, NM 87185}

\begin{abstract}
To prevent the flocculation and phase separation of nanoparticles in solution, 
nanoparticles are often functionalized with short chain surfactants. Here we 
present fully-atomistic molecular dynamics simulations which characterize how 
these functional coatings affect the interactions between nanoparticles and 
with the surrounding solvent. For 5\,nm diameter silica nanoparticles coated 
with poly(ethylene oxide) (PEO) oligomers in water, we determined the 
hydrodynamic drag on two approaching nanoparticles moving through solvent and on a single 
nanoparticle as it approaches a planar surface. In most circumstances, macroscale fluid 
theory accurately predicts the drag on these nano-scale particles.  Good agreement is seen 
with Brenner's analytical solutions for wall separations larger than the soft 
nanoparticle radius.  For two approaching coated nanoparticles, the solvent-mediated 
(velocity-independent) and lubrication (velocity-dependent) forces are purely repulsive 
and do not exhibit force oscillations that are typical of uncoated rigid spheres.
\end{abstract}

\date{\today}

\maketitle

There is increasing interest in using nanoparticles in commercial and 
industrial applications. However, effective processing of nanoparticles requires that 
they do not aggregate and often involves solvation in a fluid. 
Functionalizing the nanoparticles accomplishes both goals. The behavior of 
functionalized nanoparticles depends strongly on the attached groups. The 
behavior of bare, nonfunctionalized nanoparticles has been studied via 
experiments, theory, and simulation, as have the interactions of polymer-grafted 
surfaces \cite{klein.96,grest.99, fredrickson.91, likos.01}.  The hydrodynamic and nanoparticle-nanoparticle 
interactions involving small functionalized nanoparticles, 
however, are harder to characterize. Experimentally, it is difficult to 
manipulate and measure forces on individual nanoparticles smaller than 
$\sim$\,100\,nm. Theoretical treatments are challenging because the 
coatings are relatively short, while the particles themselves are outside the 
large radius of curvature limit. Numerical simulations of discrete 
solvent effects on nanoparticles have been impractical until now because of 
the large systems required to avoid significant finite-size effects due to the 
long-range hydrodynamic interactions.  For single-particle diffusion, these 
corrections have been shown \cite{hasimoto.59, dunweg.93} to scale as $R/L$ where $R$ 
is the particle radius and $L$ is the simulation cell length. 

Recent studies have computed the potential of mean force for bare silica 
nanoparticles in an aqueous medium, with and without electrolytes present 
\cite{jenkins.x2}. Forces between bare colloidal nanoparticles 
have also been studied in Lennard-Jones fluids and in $n$-decane 
\cite{qin.x2}. Hydrodynamic drag influenced by approach to a plane surface has 
been studied theoretically \cite{brenner.61} and compared to simulations for 
rigid spheres \cite{vergeles.96, challa.06}. Alignment effects for amorphous 
nanoparticles have also been studied \cite{fichthorn.06}. Kim {\it et al.} 
\cite{kim.05} used molecular dynamics (MD) simulations to study the 
relaxation of a fullerene molecule coated with poly(ethylene oxide) (PEO). 
Other simulations have either relied upon 
implicit solvents \cite{marla.06} or studied nanoparticles in a vacuum 
\cite{singh.07,henz.08}.

An inherent feature of interacting functionalized nanoparticles in solution is 
the multiplicity of forces at both the atomistic and nanoparticle scales.  At 
the atomistic scale these include electrostatic, van der Waals, bond, angle, 
and torsion forces.  At the nanoparticle scale, these same forces can be 
assigned to hydrodynamic drag, lubrication, and depletion forces \cite{min.08}. 
These latter 
forces are strongly dependent on the molecular structure of the nanoparticle 
coatings and solvent as well as the nanoparticle's velocity, size, and charge 
distribution.

\begin{figure}[tb]
\epsfig{file=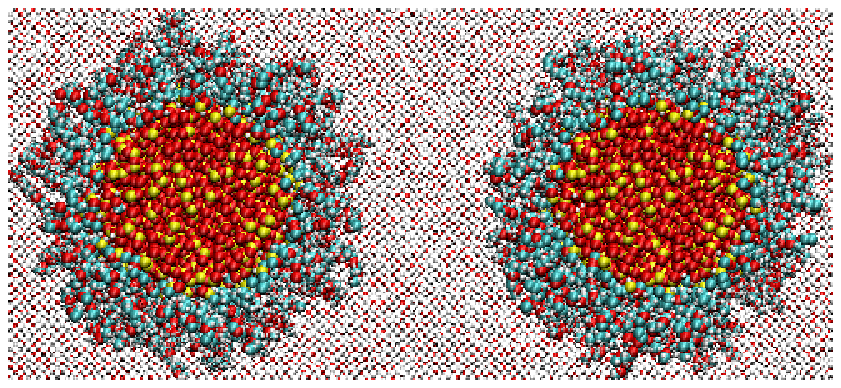,width=3in}\\
\epsfig{file=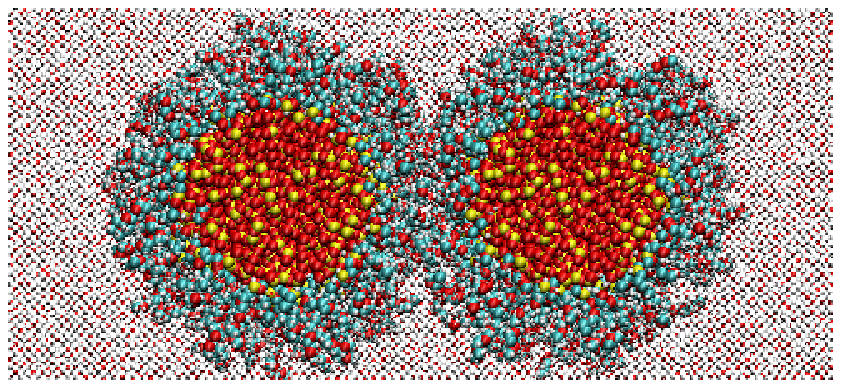,width=3in}\\
\epsfig{file=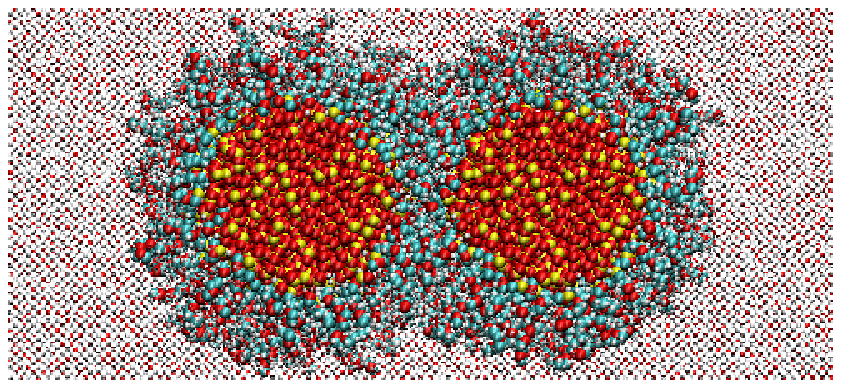,width=3in}\\
\caption{(Color online) Cross sections of silica nanoparticles coated with PEO oligomers 
in water.  Frames show a 7\,nm$\times$15\,nm view of the larger 
26\,nm$\times$26\,nm$\times$46\,nm system. The nanoparticles are shown at 
center separations of 11.5\,nm (top), 8.0\,nm (middle), and 6.0\,nm (bottom). 
Elements C, O, H, and Si are blue, red, white and yellow, respectively.}
\label{f:figure1}
\end{figure}

In this Rapid Communication we present results of MD simulations of coated nanoparticles in an explicit 
solvent. We model the forces between atoms and directly measure the resulting 
forces on nanoparticles from their motion through the discrete fluid.  To our knowledge, these 
are the first reported simulations of coated nanoparticle dynamics in an explicit, 
atomistic solvent.  These and future results will allow for more accurate and efficient 
coarse-grained potentials which characterize the interactions of small 
functionalized nanoparticles in flow environments.

Two geometries are presented.  First, a single functionalized nanoparticle approaching a 
fixed wall; and second, two nanoparticles approaching each other.  We find good agreement 
with continuum theory in the first case even for these nanoscale particles. Comparison with 
the solved hydrodynamic problem of a rigid sphere approaching a wall \cite{brenner.61} 
allows us to study system size effects. Static forces and velocity-dependent forces 
are reported.

We modeled 5\,nm diameter rigid amorphous silica nanoparticles onto which a 
passivating coating of methyl-terminated PEO oligomer chains 
(Si(OH)$_3$CH$_2$(CH$_2$CH$_2$O)$_6$(CH$_3$)) is attached. An example is shown in Figure 
\ref{f:figure1}.  The chains were chemisorbed with trisilanol groups. The grafting density 
was 3.1 chains per nm$^2$, which gave approximately 240 chains per 5 nm nanoparticle. This 
density is consistent with experimental measurements \cite{maitra.03}. We used all-atom 
force fields developed by Smith {\em et al.} for both the PEO \cite{smith.02} and the 
silica interactions \cite{smith.07}.  We used the TIP4P water model \cite{jorgensen.83}. 
While PEO-water interactions were explicitly provided, the silica-water interactions 
were determined by fitting the attractive portion of the Buckingham potentials to a 
Lennard-Jones potential and then combining them with the TIP4P potential using 
Lorentz-Berthelot mixing rules.  Simulations were carried out using the LAMMPS classical 
MD code \cite{plimpton.95}. Numerical integration was performed using the velocity-Verlet 
algorithm with 1\,fs timestep.  Long-range Coulomb interactions were calculated using the 
PPPM method \cite{hockney.88}. Unless otherwise stated, runs were made in the $NVT$ 
ensemble with Nose-Hoover thermostat at 300\,K with 100\,fs damping time.  Several were also 
run in the microcanonical ($NVE$) ensemble to test the thermostat influence.  As described 
below, no temperature drift was observed over the length of the $NVE$ runs and the $NVT$ 
and $NVE$ ensembles give statistically identical results.  However, the addition of a 
Langevin thermostat increased the drag force on the particles.  

The nanoparticle cores were cut from bulk amorphous silica and then annealed 
to produce a surface OH concentration consistent with experimental values.
The bulk silica was generated from a melt-quench process similar to the method 
of Lorenz {\it et al.} \cite{lorenz.05b}.  The nanoparticle core was treated as 
a rigid body. To build the composite systems containing water and a 
nanoparticle, we first equilibrated a large rectangular cell of TIP4P water at 
300\,K for 1\,ns. The composite system, created by inserting the 
nanoparticle into a spherical hole cut in the periodic bulk solvent, 
was then equilibrated in an $NPT$ ensemble at 300\,K and 
1\,atm pressure for 0.5\,ns. The viscosity of pure water at 1\, atm was 
calculated using the M\"uller-Plathe \cite{bordat.02} reverse perturbation 
method. The shear simulations were run on a 13.0\,nm$\times$13.0\,nm$\times$11.5\,nm 
shear cell simulation for more than 2\,ns.  The calculated viscosity of 0.55 cP 
is in agreement with previous simulations \cite{bertolini.95}.

Three single-nanoparticle systems of varying sizes were built to determine any finite-size 
effects on the viscous drag.  We varied $L_{\bot}=L_x=L_y$ perpendicular to the direction
of motion, while fixing $L_z=23.0$\,nm. For $L_{\bot}=13.0$, 26.0 and 39.0\,nm the systems 
contained approximately 400,000, 1.6 million and 3.6 million atoms, respectively.  Systems 
with two nanoparticles were created by replicating the cell in the $z$-direction. 
These systems had dimensions $L_{\bot}\times L_{\bot} \times$\ 46.0\,nm. For the largest 
simulations of more than 7 million atoms, 1\,ns of simulation time took 140 hours 
on 1024 Intel Xeon processors.

We measured the forces on the nanoparticles as a function of separation for a nanoparticle 
approaching a plane surface and for the collision of two nanoparticles. In both scenarios, 
the nanoparticle core move at constant center-of-mass velocity, which was controlled by 
displacing the rigid silica core at a fixed rate while allowing rotation. The coating and 
solvent responded dynamically. Because the displacement of the nanoparticles was 
constrained, we were able to measure the force response as a function of position for 
several approach velocities.  The force on a nanoparticle core was computed by summing all 
individual forces acting on all atoms within the core.  Forces were calculated every time 
step and averaged over 10 ps periods to remove atom-scale fluctuations.

\noindent{\bf Nanoparticle approaching a plane surface} --- The increased drag due to 
solvent-mediated interaction on a single spherical particle approaching a wall was derived 
analytically by Brenner \cite{brenner.61}. Stokes drag, valid in the asymptotic 
limit far from the wall, is modified by a multiplicative correction, $F_{\rm drag} = 6 \pi \mu R v \lambda$,
where
\begin{eqnarray*}
\lambda &=& \frac{4}{3} \left( \sinh\alpha \right) \sum_{n=1}^\infty \frac{n(n+1)}{(2n - 1)(2n + 3)} \times \\
&~&\left[ \frac{2 \sinh((2n+1)\alpha) + (2n+1)\sinh 2\alpha}{4 \sinh^2((n+\frac{1}{2})\alpha) - (2n+1)^2 \sinh^2 \alpha} - 1 \right],
\end{eqnarray*}
$\alpha = \cosh^{-1}(h/R)$, $\mu$ is the solvent bulk viscosity, $R$ is 
the radius, $v$ is the sphere velocity, and $h$ is the distance to the 
wall.

Recent explicit solvent simulations of rigid nanoparticles \cite{challa.06} have 
shown force oscillations near rigid walls. We find no oscillations for coated 
nanoparticles, while for bare silica nanoparticles we did observe oscillations.  

\begin{figure}[tb]
\epsfig{file=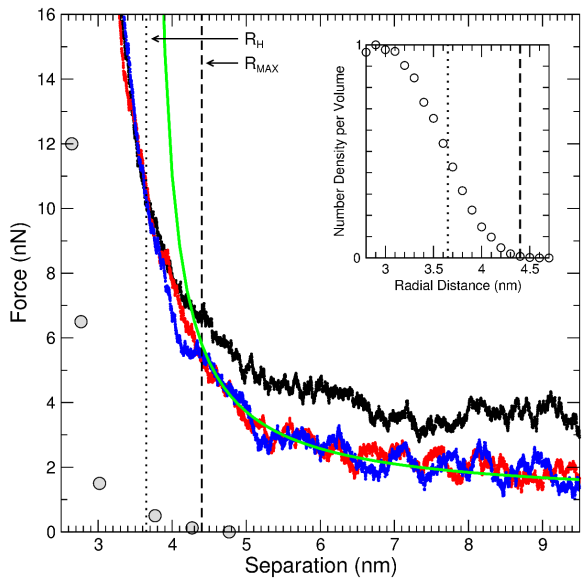,width=3in}
\caption{(Color online) Force on nanoparticles approaching a plane surface of immobile water 
at 25\,m/s as a function of center separation. The smallest system, with $L_{\bot}=$\ 13\,nm 
(black), shows a finite system size effect.  The larger systems, with 
$L_{\bot}=$\ 26\,nm (red) and $L_{\bot}=$\ 39\,nm (blue), converge to the 
Brenner prediction (green) for separations greater than 4.4\,nm.  The open circles 
are equilibrated forces in which the particles are held fixed at a given separation.  $R_{\rm max}$ 
(dashed) and $R_h$ (dotted) are defined in the text.  Inset shows the 
time-averaged coating density versus radial distance.}
\label{f:figure2}
\end{figure}

Figure \ref{f:figure2} shows the force on a PEO coated silica nanoparticle 
moving at 25\,m/s as it approaches a wall of immobile water at 
the system boundary for three system sizes.  This simple wall was selected to provide
a hydrodynamic obstacle without altering the long-range coulombic interactions of the periodic 
cell, or introducing an effective wall potential.   Two average radii are plotted as 
vertical lines: $R_{\rm max}=4.4$\,nm is the radius at 
which the particle density of the coating falls to zero, and $R_h=3.65$\,nm is the 
radius at which the particle density of the coating drops to half-maximum value. 
$R_h$ is used for the effective hydrodynamic radius in subsequent drag calculations.  The 
systems have maximum Reynolds number 0.33, indicating laminar flow and Schmidt number 180, 
indicating a momentum transfer dominated regime.

Using multiple system sizes allows us to determine the finite system-size 
effects on the measured forces. For separations greater than $R_{\rm max}$, 
finite system-size effects are evident in the smallest system. Results 
for the two larger systems show no system-size dependence and appear to have 
converged to the Brenner result.  The observed forces depart from the Brenner solution 
for distances less than $R_{\rm max}$.  The coating compresses on the surface to produce 
a softer force response than Brenner predicts for a hard sphere.  At these separations, 
short-range contact forces rather than hydrodynamic forces dominate the nanoparticle interaction.
Water molecules become trapped in the chains rather than flowing out of the gap.
Moreover, the direct van der Waal and Coulomb interations between particle coating and wall becomes significant.
  Nanoparticles in all three system sizes 
feel identical forces for separations less than $R_h$ at a given velocity.  For 
separations greater than $R_{\rm max}$, the lubrication forces are strongly dependent on 
system size for small systems ($L/R \sim 3$), but fall away rapidly with increase system 
size.  This perhaps indicates that the short-range hydrodynamics dominate over long-range 
hydrodynamics which are predicted to fall away much more slowly, scaling as $R/L$ 
\cite{dunweg.93, hasimoto.59}.

\noindent{\bf Force between two nanoparticles} ---
The cross sections in Fig.~\ref{f:figure1} depict the system we used to 
study the forces between two nanoparticles.  The two-nanoparticle dynamic simulations 
were run with system size $L_{\bot}=$\ 26\,nm and $L_z=$\ 46\,nm.  Equilibration runs
for smaller $L_{\bot}=$\ 13\,nm and $L_z=$\ 23\,nm systems are also reported.  As seen in 
Fig.~\ref{f:figure1}, the chains begin to touch when the center separation is 
approximately 8.5\,nm. At 6.0\,nm, the tethered chains are highly compressed.

Relative velocities of 5 and 50 m/s were used to determine the velocity-dependent 
lubrication forces.  Starting at a core-core separation of 23 nm the
nanoparticles were brought together until the forces between the two nanoparticles 
diverged. For the velocity independent solvation forces, the cores were held at various 
fixed positions taken from the finite velocity approach runs and each allowed to relax 
for 2 ns.

\begin{figure}[tb]
\epsfig{file=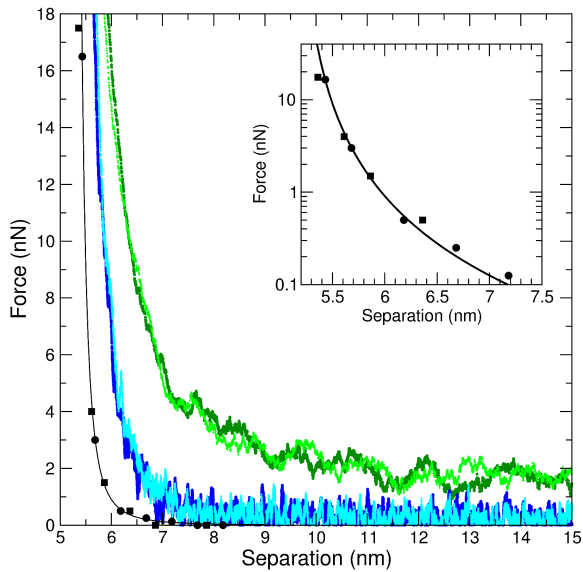,width=3in}
\caption{(color online) Force on approaching nanoparticles with relative velocities of 5 m/s 
(light blue and dark blue) and 50 m/s (light green and dark green) as a function of center separation.
The lighter colors are data from runs in the NVE ensemble and the darker colors 
were run in the thermostated NVT ensemble.  Data for equilibrated configurations are shown 
in black. Inset shows the equilibrated data as a log-linear plot. 
Circles and squares are for $L_{\bot}$ of 13\,nm and 26\,nm, respectively.} 
\label{f:figure3}
\end{figure}

Figure~\ref{f:figure3} shows the force measured as a function of separation. The 
$NVT$ and $NVE$ ensembles give statistically identical results.  The force is purely 
repulsive and, although fluctuating, on average increases monotonically with decreasing 
separation. The force response is consistent with the theoretical linear scaling in 
velocity and goes asymptotically to the single-particle Stokes drag in the limit of 
large separation.
As in the single-particle study, we do not observe oscillations in the force 
which have been observed with two rigid sphere nanoparticles in explicit 
solvent \cite{fichthorn.06}.

Finite-system effects are expected for small systems at higher velocities. 
However, the equilibrated solvation forces showed no system size 
effects even for the smallest systems.  The inset in figure \ref{f:figure3} 
shows equilibrated forces computed in $L_{\bot}=$\ 13\,nm and 26\,nm systems.
The equilibrated force is plotted at fixed separations and fit within this range
of separation by  
$F_{\rm equil} \propto (1/r)^{2.5}$, where $r$ is the core surface separation.  
%The effective osmotic virial coefficient, $\Gamma_2= -2 \pi \int_0^\infty [1-e^{-U(r)/kT}]r^2dr$, 
%is 100. $U(r)$ was determined analytically from $F_{\rm equil}$

The two-nanoparticle data can be broken into three regimes, which we will refer to as large 
separations (greater than 8\,nm), moderate separations (between 6 and 8\,nm), and very small 
separations (less than 6\,nm). These regimes are characterized by nanoparticle-nanoparticle 
forces which are non-interacting, weakly interacting, and strongly interacting, 
respectively. At large separations, the force on the nanoparticles is dominated by a 
constant resistance to motion by the solvent, and the system is clearly in the viscous drag 
regime. The force on the nanoparticle in this regime is essentially constant as a function 
of separation. The magnitude of the force scales linearly with velocity as hydrodynamic 
theory predicts in this velocity range.
At moderate separation distances, the separation-dependent interaction of the 
coatings dominates the force.  Both coating interactions and hydrodynamic forces 
play significant roles.  The hydrodynamic forces increase with proximity as the 
solvent is forced from the small space between the nanoparticles. We note that in 
this region the force is strongly separation- and velocity-dependent. Finally, for 
very small separations, the chains of the coating become close-packed and the 
force increases rapidly. This final region is unlikely to be 
explored by nanoparticles in solution as the energies associated with these 
separations are orders of magnitude above the thermal background $k_BT$ under 
normal processing conditions.

In conclusion, we have presented results for large-scale simulation of coated 
nanoparticles in an explicit solvent.  We measured the forces on 
these nanoparticles as a function of separation in two geometries and for 
several velocities. For separations larger than $R_{\rm max}$, 
analytical hydrodynamic results for colloidal particles held even at the 
nanoscale.  For two nanoparticles, at large separations the 
forces were dominated by hydrodynamic forces, specifically viscous drag.
For separations less than $2R_{\rm max}$, the interparticle forces were 
dominated by contact and short-range hydrodynamic forces.  In all cases, 
the functional coating completely suppressed the force oscillations which are
observed for uncoated rigid particles of this size in explicit solvents.  
We observed a similar qualitative response for alkane-coated silica 
nanoparticles in decane.  Results from 
an extensive series of simulations which investigate the role of chain length, 
grafting density, and core nanoparticle size will be reported elsewhere. 

The authors thank Frank van Swol and Burkhard D\"unweg for useful discussions.
We thank the New Mexico Computing Application Center (NMCAC) for generous 
allocation of computer time.  This work is supported by the Laboratory 
Directed Research and Development program at Sandia National Laboratories. 
Sandia is a multiprogram laboratory operated by Sandia Corporation, a 
Lockheed Martin Company, for the United States Department of Energy's 
National Nuclear Security Administration under Contract DE-AC04-94AL85000.

\bibliographystyle{apsrev}
%\bibliography{nano2}

\end{document}